   \documentclass[final,5p,times,twocolumn]{elsarticle}

\usepackage{amssymb}
\usepackage{amsmath}

\usepackage{verbatim}

\usepackage{lineno}

\journal{Nuclear Instruments and Methods in Physics Research A}

\begin{document}

\begin{frontmatter}



\title{Comment on "Monochromatization interaction region optics design for direct s-channel Higgs production at FCC-ee"}


\author{D. Shatilov} 
\ead{dshatilov@niu.edu}

\affiliation{organization={Northern Illinois University}, 
            addressline={1425 W. Lincoln Hwy.}, 
            city={DeKalb},
            postcode={60115}, 
            state={IL},
            country={USA}}

\begin{abstract}
The original article \cite{Zhang2025} can be logically divided into two parts: 1) the selection of main parameters for monochromatization and 2) interaction region optics design; the comment pertains only to the first part. The authors of \cite{Zhang2025} state that "The purpose of this paper is to report on the development of realistic IR optics designs for monochromatization at the FCC-ee". However, the proposed parameters do not seem very realistic and raise many questions; due to space limitations, we will only consider the most important ones.
\end{abstract}


\end{frontmatter}

\section{Piwinski angle and collision overlap area}
The Piwinski angle is not actually an angle, but a dimensionless parameter. Therefore, the indication that it is measured in radians (Tables 1 and 2) is inappropriate.

In the general case, the length of interaction area \cite{Shatilov2018} (that is Ref. [14] in \cite{Zhang2025}) is expressed as
\begin{equation}
L_i = \frac{\sigma_z}{\sqrt{1+\varphi^2}} .
\label{eq:L_i}
\end{equation}
For $\varphi \gg 1$ and $\theta_c \ll 1$, this can be simplified to:
\begin{equation}
L_i \approx \frac{\sigma_z}{\varphi} \approx \frac{2 \sigma_x}{\theta_c} .
\label{eq:L_i2}
\end{equation}
As the authors rightly noted, the $L_i \big/ \beta_y^*$ ratio characterizes the hourglass effect. But using Eq. (\ref{eq:L_i2}) for collision overlap area in \cite{Zhang2025} is unjustified, since the typical value of Piwinski angle in Tables 1 and 2 (with monochromatization) is $\varphi \approx 1$. The exact expression (\ref{eq:L_i}) for $L_i$ should be used instead. Accordingly, the line "Overlap collision ..." in the Tables should be corrected, which will also affect $R_{hg}$ and  the calculated luminosity.

\section{Synchrotron tune and $t\bar{t}$ optics}
The value and success of the direct s-channel Higgs production experiment depend significantly on the accuracy of the centre-of-mass energy determination, which requires precise beam energy calibration. The most accurate method that everyone is counting on is based on measuring the spin precession frequency. These measurements are performed using resonant depolarization, which in turn imposes stringent requirements on the synchrotron tune.

The modulation of the spin precession frequency by synchrotron oscillations \cite{Polar2019} is characterized by the synchrotron modulation index: $\zeta = \nu_0 \sigma_{\delta} \big/ Q_s$, where $\nu_0 \approx 142$ is the spin tune at 62.5 GeV and $\sigma_{\delta} \approx 5.5 \cdot 10^{-4}$. For a reliable determination of the spin frequency, one should aim for $\zeta < 1.5$, and in any case $\zeta < 2$ must be satisfied. It follows that $Q_s$ should be at least greater than 0.04, and preferably greater than 0.05. In principle, this is possible in $t\bar{t}$ optics, but it will require raising the RF voltage from 0.17 to $1.0-1.2$ GV. As a result, $\sigma_z$ will decrease several times, which will have both positive and negative consequences. Positive aspects include weakening the hourglass effect and reducing the Piwinski angle, which is necessary for monochromatization. Negative aspects include enhancement of collective instabilities for short bunches and increased beamstrahlung, which will lead to a significant increase in $\varepsilon_x$ and a decrease in $\lambda$. To avoid this, it will be necessary to reduce the bunch population several times (and, consequently, increase the number of bunches), which will lead to a decrease in luminosity. In any case, many of the values in Table 2 will have to change significantly, so in its current form, this Table does not make much sense.

\section{Vertical dispersion}
The time available for direct s-channel Higgs production experiment will be limited. Therefore, it is critical to have a sufficiently high luminosity to obtain the required number of events. The dependence of the Higgs event rate on luminosity and $\sigma_W$ can be expressed as
\begin{equation}
\dot{N}_H \propto \frac{L}{\sqrt{\Gamma_H^2 + \sigma_W^2}} ,
\label{eq:NH}
\end{equation}
where $\Gamma_H \approx 4.1$ MeV. The original idea was to increase $\dot{N}_H$ by monochromatization in the hope that the reduction in the luminosity would be small. Then the goals became more realistic: $L$ can decrease in approximately the same proportion as $\sigma_W$, so that $\dot{N}_H$ remains almost unchanged. But if monochromatization leads to a noticeable (several times) decrease in $\dot{N}_H$, then it loses its meaning.

With this in mind, let us consider the scheme with $D_y^* \neq 0$ again. On the one hand, it is not hindered by the horizontal crossing angle, and this is the main advantage. On the other hand, increasing $\sigma_y^*$ leads to a decrease in $L$ by a factor of $\lambda$, which is still within acceptable limits. But this works only without BS, that is, for a small bunch population and, consequently, at low luminosity. And taking BS into account, $\dot{N}_H$ will inevitably decrease many times over. That is why the option with vertical dispersion was not even considered in the CDR \cite{FCC_CDR2019}.

The advantage of horizontal dispersion over vertical one becomes clear if we take into account the fact that for flat beams, increasing $\sigma_x^*$ weakens BS, while increasing $\sigma_y^*$ does not. The results presented in \cite{Zhang2025} also demonstrate a multiple reduction in $\dot{N}_H$ in the case of $D_y^* \neq 0$. Therefore, to avoid misleading readers, it would be useful to make a clear and distinct conclusion that such a scheme should be rejected.

\section{Monochromatization factor}
Eq. (3) in \cite{Zhang2025} is obviously incorrect. To understand this, it suffices to consider the limiting case $D_y^* = 0$, $\varphi \gg 1$ and $D_x^* \sigma_{\delta} \big/ \sqrt{\varepsilon_x \beta_x^*} \gg \varphi$. Then we get $\lambda \gg 1$, which cannot be the case for $\varphi \gg 1$ -- this is clear from the geometry shown in Fig. 2 in \cite{Zhang2025}. Everything is explained simply: the authors took Eq. (3) from \cite{Bogom2017} (that is Ref. [13] in \cite{Zhang2025}), see Eq. (17) there. However, they did not take into account that the Piwinski angle is defined incorrectly in \cite{Bogom2017}: instead of the full horizontal beam size $\sigma_x^*$, the betatron size $\sigma_{x \beta}^* = \sqrt{\varepsilon_x \beta_x^*}$ was substituted into the formula for determining $\varphi$; this issue is considered in detail in \cite{Shatilov2018}.

The correct expression for $\lambda$ in the case $D_y^* = 0$ and without taking into account the hourglass effect can be found in \cite{Shatilov2018, FCC_CDR2019}:
\begin{equation}
\lambda = \sqrt{1 + \frac{\lambda_m^2}{1 + \varphi^2(1+\lambda_m^2)}} ,
\label{eq:lambda}
\end{equation}
where $\lambda_m$ is the ratio between synchrotron and betatron horizontal beam sizes:
\begin{equation}
\lambda_m = \frac{D_x^* \sigma_{\delta}}{\sqrt{\varepsilon_x \beta_x^*}} .
\label{eq:lambda_m}
\end{equation}
From Eq. (\ref{eq:lambda}) it follows that $\sigma_W$ should increase from 26.4 to 37.9 MeV for ZH4IP and from 20.2 to 41.8 MeV for ZH2IP, making the latter scheme less attractive.

Given that the monochromatization occurs in the horizontal plane, it might seem that only the horizontal hourglass (which is negligible, since $\beta_x^* \gg \sigma_z$) matters. However, the vertical hourglass will also affect $\lambda$, as it leads to a change in the luminosity distribution along the azimuth, which is in some sense equivalent to a reduction in $\sigma_z$. Accordingly, the "effective" Piwinski angle will decrease, and $\lambda$ will increase. To evaluate the magnitude of this effect, we used modeling with the Lifetrac beam-beam code \cite{Lifetrac1996}. For comparison purposes, exactly the same initial conditions were used as in \cite{Zhang2025}: BS was switched off, and the beams were represented as an ideal Gaussian distribution defined by the global optical parameters taken from Table 1 (with BS). In simulations, the vertical hourglass can be reduced to almost zero by multiple increase in $\beta_y^*$, and in this case, excellent agreement with Eq. (\ref{eq:lambda}) was obtained. With a nominal $\beta_y^* = 1$ mm, we obtained $\sigma_W$ equal to 31.3 MeV for ZH4IP and 32 MeV for ZH2IP. These values do not agree well with the simulation results presented in Table 3, which raises questions about this simulation.

As can be seen, the positive effect of the vertical hourglass on $\lambda$ can be quite significant, but it weakens as $\varphi$ decreases, which is necessary for monochromatization. Besides, large hourglass would be impractical for a number of other reasons. And for $\beta_y^* \approx L_i$, Eq. (\ref{eq:lambda}) can be quite useful for estimations. It should also be noted that the precise expression for $\lambda$, taking into account the hourglass effect, is not particularly useful in our case, since there are no exact analytical formulas for calculating $\varepsilon_x$ considering the BS effect. The equations for $\varepsilon_x$ in \cite{Zhang2025} are suitable for a rough estimate, while realistic values can only be obtained through simulations, where the values of $\sigma_W$ and $\lambda$ can also be found directly.

\section{Vertical emittance}
In Section 3.3, the authors reasonably noted that $\varepsilon_y$ was calculated with the assumed (initial) FCC-ee coupling ratio $\varepsilon_{y0} \big/ \varepsilon_{x0} = 0.2 \%$. If $D_x^* \neq 0$, $\varepsilon_x$ will increase several times due to BS, and it would be too optimistic to assume that $\varepsilon_y$ will remain unchanged.

The excitation of $\varepsilon_y$ depends on two factors: 1) betatron coupling and 2) vertical dispersion (mainly in the arcs) which also arises due to coupling. An increase in $\varepsilon_x$ due to BS will lead to an increase in $\varepsilon_y$ due to the first factor, but not the second. To evaluate this effect, it is necessary to know the ratio of the contributions of these two factors to $\varepsilon_{y0}$. However, this depends on the specific sources of X--Y coupling in the realistic magnetic lattice with errors, misalignments, etc., as well as on the methods used to correct them, and all these details are unknown for now. For a realistic assessment, it is best to use a set of simulations (with different random seeds) that take into account all the details mentioned above. If such a model is not available, then only approximate assumptions can be made.

For example, $\varepsilon_x$ increased about 3.7 times for ZH4IP. Based on the conservative assumption that the contribution of the first factor to the nominal $\varepsilon_{y0}$ is at least one-third, we will get a 1.9-fold increase in $\varepsilon_y$ due to BS. Accordingly, the luminosity will decrease by approximately 1.4 times, and this factor (with appropriate explanations) should be taken into account in the Table of parameters to make it more realistic. 

\section{Parameter optimization}
The Abstract of [1] contains a phrase that is perceived as a promise that will be fulfilled: "This measurement is significantly facilitated if the CM energy spread of $e^+e^-$ collisions can be reduced to a level comparable to the natural width of the Higgs boson, $\Gamma_H$ = 4.1 MeV, without substantial loss in luminosity". And here is the result: $\sigma_W$ became 31.3 MeV, while the luminosity decreased by $\sim$2.1 times (taking into account the aforementioned factor of 1.4). Is this what we were aiming for?

As can be seen from Eq. (\ref{eq:lambda}), to get $\lambda \gg 1$ in a collision scheme with crossing angle and without crab crossing, we need to make $\varphi \ll 1$. To achieve this, $\sigma_z$ can be reduced (for example, $\sim$1.5 times) by increasing the RF voltage and $D_x^*$ should be significantly increased. The statement about a "minimum bunch spacing of 25 ns" also raises questions. Perhaps this limitation exists at low energy (Z-pole, 45.6 GeV), where the total beam current is 3.5 times higher and the damping is 2.6 times weaker. However, at 62.5 GeV things become easier, so decreasing $N_b$ (along with increasing the number of bunches $n_b$) can be a useful option.

An adequate example of parameter optimization can be found in the CDR \cite{FCC_CDR2019}, where the emphasis was on achieving the required $\sigma_W$ and luminosity values. Of course, much has changed since then, including the perimeter, the number of IPs, the arc cell type at low energies and hence $\alpha_C$, $\varepsilon_x$, etc. Consequently, it will be necessary to re-optimize all the parameters, and then, which will be the most difficult part, implement this in the IR optics design.

\section*{Declaration of competing interest}
The authors declare that they have no known competing financial interests or personal relationships that could have appeared to influence the work reported in this paper.

\section*{Acknowledgements}
This manuscript has been authored by Fermi Forward Discovery Group, LLC under Contract No. 89243024CSC000002 with the U.S. Department of Energy, Office of Science, Office of High Energy Physics. This work used resources of the Center for Research Computing and Data at Northern Illinois University.

\section*{Data availability}
Data will be made available on request.




\end{document}